\documentclass[twocolumn,aps,prl,nofootinbib]{revtex4-2}
\usepackage{natbib}
\usepackage{subcaption}

\usepackage{graphicx}
\usepackage{amsmath}
\usepackage[nopatch]{microtype}
\usepackage{booktabs}
\usepackage{hyperref}
\usepackage{url}
\usepackage{graphics}
\usepackage{xspace}
\usepackage{tikz}
\usepackage{amsmath,amssymb}
\usepackage{bm}
\usepackage{xcolor} 
\usepackage{soul}
\newcommand{\hcon}{\ensuremath{\mathrm{km~s^{-1}~Mpc^{-1}}}\xspace}
\usepackage{graphicx}
\usepackage{dsfont}
\usepackage{dcolumn}
\usepackage{bm}

\begin{document}

\preprint{APS/123-QED}

\title{Hostbusters: constraining $H_0$ and binary formation channels\\
through gravitational-wave host galaxy associations}

\author{Madeline L. Cross-Parkin}
\email{[m.crossparkin@uq.edu.au]}
\affiliation{%
School of Mathematics and Physics, The University of Queensland, Brisbane, Queensland 4072, Australia}%

\author{Ignacio Magaña Hernandez}
\affiliation{%
McWilliams Center for Cosmology and Astrophysics,
Department of Physics,
Carnegie Mellon University,
Pittsburgh, Pennsylvania 15213, USA}%

\author{Antonella Palmese}
\affiliation{%
McWilliams Center for Cosmology and Astrophysics,
Department of Physics,
Carnegie Mellon University,
Pittsburgh, Pennsylvania 15213, USA}%
    
\author{Paul D. Lasky}
\affiliation{%
School of Physics and Astronomy, Monash University, Clayton, Victoria 3800, Australia}%
\affiliation{%
OzGrav: The ARC Centre of Excellence for Gravitational-Wave Discovery, Clayton, Victoria 3800, Australia}%

\author{Eric Thrane}
\affiliation{%
School of Physics and Astronomy, Monash University, Clayton, Victoria 3800, Australia}%
\affiliation{%
OzGrav: The ARC Centre of Excellence for Gravitational-Wave Discovery, Clayton, Victoria 3800, Australia}%

\begin{abstract}
Gravitational-wave dark sirens can be combined with galaxy catalogues to provide a measurement of the Hubble constant. This inference requires assigning each potential host galaxy a probability of hosting a merger, but the appropriate weighting is uncertain because the binary merger rate may trace different galaxy properties, such as stellar mass or star formation rate, depending on the underlying formation channel. We introduce a flexible framework in which the dark-siren population is described as a mixture of two host-weighting prescriptions, with the mixing fraction inferred jointly with $H_0$. These components may be interpreted as subpopulations associated with distinct binary-formation environments. Using simulated catalogues, we show that marginalising over uncertainty in the host-population mixture preserves unbiased constraints on $H_0$, while allowing the gravitational-wave data to constrain the relative contribution of the two host populations. In particular, an A\#-like detector network with $\mathcal{O}(10^3)$ dark sirens may be able to exclude scenarios in which the host population traces only one of the two weighting prescriptions. Dark-siren analyses may therefore provide robust cosmological measurements and new information about how merging compact binaries trace the galaxy population.
\end{abstract}

\maketitle

\section{Introduction}
Measuring the present-day expansion rate of the Universe is a central goal of modern cosmology, particularly in light of the $\sim5\sigma$ tension \citep{Verde:2019ivm,CosmoVerseNetwork:2025alb,Abdalla:2022yfr} between late-Universe distance-ladder measurements \citep{Riess:2021jrx} and early-Universe constraints inferred from the cosmic microwave background \citep{Planck:2018vyg}. This discrepancy has motivated measurements of the Hubble constant, $H_0$, that rely on different observables and which have different sources of uncertainty. 

Gravitational-wave standard sirens provide a direct probe of cosmology through the measurement of luminosity distances to compact binary mergers. For binary black hole mergers without electromagnetic counterparts, inferring $H_0$ relies on the statistical association between gravitational-wave events and galaxies in large-scale structure catalogues. Several approaches for measuring $H_0$ with such dark sirens have been developed to exploit this information \cite{2026enap....5..557P, Cheng:2026atn, crossparkin2026darksirencrosscorrelationssensitivity, Mukherjee_2021,Mukherjee_2024}. The most common is the galaxy catalogue approach, in which candidate host galaxies are weighted by how likely they are to host the event, and a posterior on $H_0$ is obtained by marginalising over all potential hosts \citep{Chen:2017rfc, Finke:2021aom, 2023AJ....166...22G, Gray:2019ksv, Holz_2005, LIGOScientific:2018gmd, LIGOScientific:2021aug, PhysRevD.86.043011, Soares_Santos_2019}.


An important aspect of this method is the choice of how candidate host galaxies are weighted. Many dark siren analyses assume that all galaxies are equally likely to host a gravitational wave event, but, as emphasised by \citet{Hanselman_2025}, uniform weighting is itself an astrophysical assumption rather than an assumption-free baseline. 

Binary merger rates are expected to depend on host-galaxy properties such as stellar mass, luminosity, star-formation history, and redshift \citep{Artale_2019a, Artale_2019b}, motivating the use of physically informed weighting prescriptions that more closely reflect compact-binary formation. The choice of how to weight different host galaxies affects our inferences about $H_0$; incorrect assumptions about the host population have been shown to bias dark siren measurements of $H_0$ \citep{Perna_2025,Hanselman_2025,li2025usinggravitationalwavedark,alfradique2025systematicbiasdarksiren}. While this bias may be reduced when the galaxy catalogue is sufficiently informative \citep{dang2026yearnight1mathbfh0,Palmese_2023},
this raises a natural question: if different plausible weighting prescriptions lead to different cosmological constraints, how do we determine which one best describes the true merger population?

To address this, we adopt a more flexible framework in which the data is used to infer the relative contribution of different host weighting prescriptions. In doing so, the host model is no longer treated solely as a source of systematic uncertainty: constraining it can both protect the inference of $H_0$ from bias and provide information about the environments in which compact binaries form and merge.

In this work, we model the host probability as a mixture of two astrophysically motivated tracers: $r$-band luminosity and $g-r$ colour. The $r$-band luminosity approximately traces stellar mass and therefore older, accumulated stellar populations, motivating its use for long-lived merger environments such as globular clusters \citep{Rodriguez_2016}. In contrast, $g-r$ colour is sensitive to stellar age and recent star formation, making it a useful tracer for more actively star-forming populations and some isolated field-binary pathways \citep{Di_Carlo_2019}. We therefore use these as representative mass- and star-formation-weighted host prescriptions, respectively.

By allowing the contribution of these two prescriptions to vary, we demonstrate that the same framework can simultaneously (1) recover unbiased estimates of $H_0$ and (2) constrain the relative contribution of merger populations associated with different host-galaxy environments. Importantly, the statistical framework itself is not specific to $r$-band luminosity and colour. These choices provide a physically motivated demonstration, but the likelihood can be readily adapted to alternative galaxy properties or weighting prescriptions if other observables provide more informative tracers of compact-binary formation.

\textcolor{black}{
While this manuscript was in preparation, we became aware of two related studies
\citep{gray2026expandingscopedarksiren, agapito2026gravitationalwavedarksirensastrophysical}. Both consider host-weighting models in which the dependence of the host probability on galaxy properties is constrained jointly with the cosmological parameters. We provide a comparison between these approaches and the framework developed here in Section~\ref{sec: comparison}.}

This work is organised as follows. In Section~\ref{sec:method} we outline the statistical framework used in our analysis, as well as the process for generating gravitational wave events, before applying it to the simulated gravitational-wave catalogues described in Sec.~\ref{subsec:sims}. In Section~\ref{sec:results} we present the results from our analysis, and in Section~\ref{sec:disc} we report our conclusions.

\section{Methodology}
\label{sec:method}
This section first introduces the general hierarchical framework in Sec.~\ref{subsec:standard},~\ref{subsec:pop} and~\ref{subsec:post}. We then specialise this framework to the host-galaxy mixture model in Sec.~\ref{subsec:weights}.

\subsection{Bayesian framework}
\label{subsec:standard}

We employ Bayesian hierarchical inference to obtain the posterior for a set of $N_{\rm det}$ gravitational-wave events. For each individual event, the full gravitational-wave signal is described by fifteen source parameters. However, for our dark siren analysis, the relevant information is luminosity distance and sky localisation, so we work with the likelihood marginalised over the remaining source parameters. We use a Fisher approximation, in which the gravitational-wave likelihood for event $k$ is represented by a multivariate Gaussian,
\begin{equation}
\mathcal{L}
\left(
d_{L,k}^{\rm obs},
\hat{\Omega}_k^{\rm obs}
\mid
d_{L,k},
\hat{\Omega}_k
\right),
\label{eq:gwlikelihood}
\end{equation}
where $d_{L,k}^{\rm obs}$ and $\hat{\Omega}_k^{\rm obs}$ are the observed luminosity distance and sky position of gravitational wave event $k$, respectively, while $d_{L,k}$ and $\hat{\Omega}$ denote the corresponding true source parameters. We define the angular coordinates as $\hat{\Omega}=\{{\rm RA},\sin({\rm Dec})\}$. We model this as a truncated multivariate Gaussian; however, the likelihood has negligible support near the boundaries, so the effect of the truncation is minimal.

\subsection{Galaxy catalogue and population model}
\label{subsec:pop}
We assume a complete galaxy catalogue containing $n$ host galaxies,
with redshifts and sky positions
$\{z_i^{\rm obs},\hat{\Omega}_i^{\rm obs}\}_{i=1}^{n}$.
Each galaxy is assigned two weights, $w_i$ and $w_i'$, corresponding to two alternative prescriptions for how the merger population traces the galaxy catalogue. Each set of weights is normalised such that its sum over all catalogue galaxies is unity. We describe the motivation for these weights and how they are calculated in Section~\ref{subsec:weights}.

Our population model contains two hyperparameters,
$\Lambda=\{H_0,\xi\}$. The Hubble constant, $H_0$, determines the luminosity distance--redshift relation, while $\xi$ controls the relative contribution of the two host-population prescriptions. For an individual event, $\xi$ can be interpreted as the prior probability that the event follows the first host-weighting prescription. When combining many events, $\xi$ can equivalently be interpreted as the population mixing fraction between the two prescriptions.

The prior is
\begin{align}
\pi(d_L,\hat{\Omega}\mid & H_0,\xi)=
\xi
\sum_{i=1}^{n}
w_i\,
\pi\!\left(
d_L
|z_i^{\rm obs},H_0
\right)\,
\delta^{(2)}
\left(
\hat{\Omega}-\hat{\Omega}_i^{\rm obs}
\right)
\nonumber\\
&+
(1-\xi)
\sum_{i=1}^{n}
w_i'\,
\pi\!\left(
d_L
|z_i^{\rm obs},H_0
\right)\,
\delta^{(2)}
\left(
\hat{\Omega}-\hat{\Omega}_i^{\rm obs}
\right).
\label{eq:event-catalogue-prior}
\end{align}
We treat the galaxy sky positions as exact, so the angular part of the catalogue prior is represented by a delta function.
However, the redshift is not known precisely, and so $\pi(d_L|z_i^{\rm obs},H_0)$ is a smooth distribution peaking at the luminosity distance predicted by the assumed cosmology, $d_L(z_i^{\rm obs},H_0)$, with some width.

We calculate this distribution by assuming a Gaussian prior on $z_i$, centred on the observed redshift $z_i^{\rm obs}$ with width $\sigma_{z,i}=10^{-3}(1+z_i^{\rm obs})$, representative of the spectroscopic redshift precision expected from Euclid \citep{laureijs2011eucliddefinitionstudyreport}. We then transform this redshift prior into a luminosity-distance prior using the appropriate Jacobian.

Before moving on, we note that $d_{L,k}^\text{obs}$ refers to the measured gravitational-wave distances whereas $z_i^\text{obs}$ refers to the measured \textit{galaxy} redshift of \textit{galaxy} $i$.
To help keep these two things straight, we consistently use $k$ to label gravitational-wave events and $i$ to label galaxies.

\subsection{Posterior}
\label{subsec:post}
For a single gravitational-wave event, we can calculate the likelihood of the gravitational-wave data given the hyperparameters:
\begin{align}
\mathcal{L}(d_{L,k}^\text{obs}, \hat\Omega_k^\text{obs} | H_0,\xi)= & 
\int d d_L
\int d\hat{\Omega}\,
\mathcal{L}
\left(
d_{L,k}^{\rm obs},
\hat{\Omega}_k^{\rm obs}
\mid
d_{L,k},
\hat{\Omega}_k
\right) \nonumber\\
& \pi_k
\left(
d_L,\hat{\Omega}
\mid
H_0,\xi
\right).
\label{eq:catalogue-marginalised-likelihood}
\end{align}

Substituting the prior and carrying out the angular integral gives
\begin{align}
\mathcal{L}(d_{L,k}^\text{obs}, \hat\Omega_k^\text{obs} | & H_0,\xi)
=
\xi
\sum_{i=1}^{n}
w_i
\int d d_L\,
\pi(
d_L\mid z_i^{\rm obs},H_0) \nonumber\\
&\mathcal{L}
\left(
d_{L,k}^{\rm obs},
\hat{\Omega}_k^{\rm obs}
\mid
d_L,
\hat{\Omega}_i^{\rm obs}
\right)
\nonumber\\
&+
(1-\xi)
\sum_  
{i=1}^{n}
w_i'
\int d(d_L) \, 
\pi(
d_L\mid z_i^{\rm obs},H_0) \nonumber\\
& \mathcal{L}
\left(
d_{L,k}^{\rm obs},
\hat{\Omega}_k^{\rm obs}
\mid
d_L,
\hat{\Omega}_i^{\rm obs}
\right).
\label{eq:event-likelihood}
\end{align}

For a set of $N_{\rm det}$ independent gravitational-wave events, the posterior on the hyperparameters is then
\begin{align}
p
\left(
H_0,\xi
\mid
\{d_{L,k}\}_{k=1}^{N}
\right)
\propto
\pi(H_0,\xi)
\prod_{k=1}^{N}
\frac{
\mathcal{L}(d_{L,k}^\text{obs}, \hat\Omega_k^\text{obs} |H_0,\xi)
}{
p_{\rm det}(H_0,\xi)
} ,
\label{eq:hyperposterior}
\end{align}
where $p_{\rm det}(H_0,\xi)$ gives the probability that a source drawn from the population described by $(H_0,\xi)$ satisfies the detection threshold. It is given by
\begin{align} 
p_{\rm det}(H_0,\xi) = \int d d_L\, d\hat{\Omega}\, p_{\rm det} \left( d_L,\hat{\Omega} \right) p \left( d_L,\hat{\Omega} \mid H_0,\xi \right). 
\label{eq:pdet_hyper} 
\end{align} 
The source-level detection probability is 
\begin{align} 
p_{\rm det} \left( d_L,\hat{\Omega} \right) = \int_{\rho>\rho_{\rm th}} d d_L^{\rm obs}\, d\hat{\Omega}^{\rm obs}\, \mathcal{L} \left( d_L^{\rm obs}, \hat{\Omega}^{\rm obs} \mid d_L,\hat{\Omega} \right), \label{eq:pdet} 
\end{align} 
which gives the probability that a source with true parameters $(d_L,\hat{\Omega})$ satisfies the detection threshold and therefore enters the observed sample; see, e.g., Ref.~\cite{intro}.

\subsection{Weights}
\label{subsec:weights}
In Eq.~\ref{eq:event-catalogue-prior}, we employ two weights, $w_i$ and $w_i'$, which encode the relative probability that galaxy $i$ hosts a gravitational-wave event. Assigning the same weight to every galaxy ($w_i=w_i'=\mathrm{const}$) recovers the commonly adopted uniform-weighting case. However, this is unlikely to provide a realistic description of the host population. Galaxies differ substantially in properties, such as their stellar content and star-formation history, which are expected to influence their gravitational-wave merger rates. A more massive or more luminous galaxy may therefore contribute more mergers than a faint, low-mass galaxy. Such galaxies should consequently not be assigned equal host probabilities.

The less obvious question is how these differences should be used to calculate the weights $w, w'$. We consider two physically motivated prescriptions: \textit{star formation weighting} and \textit{mass weighting}. We represent these using galaxy colour and $r$-band luminosity, respectively, which trace complementary properties of the underlying stellar population.

For \textit{mass weighting}, we use the $r$-band luminosity as a proxy for stellar mass,
\begin{align}
L_{r,i} \propto 10^{-0.4M_{r,i}},
\end{align}
where $M_{r,i}$ is the absolute magnitude of galaxy $i$ in the $r$ band. A mass-weighted prescription is particularly well motivated for compact-binary populations with appreciable delay times between stellar formation and merger, for which the present-day merger rate is expected to correlate strongly with the accumulated stellar mass of the host galaxy \citep{Artale_2019b}.

For \textit{star formation weighting}, we use the galaxy colour,
\begin{align}
c_i = g_i-r_i,
\label{eq:color}
\end{align}
where $g_i$ and $r_i$ are the $g$- and $r$-band apparent magnitudes of galaxy $i$. The colour is sensitive to stellar-population properties and star-formation activity. Since the common distance modulus cancels in a colour, the apparent and absolute colours are equivalent when defined in the same rest-frame bands. We therefore use rest-frame colours in this work. For real photometric data, however, recovering rest-frame colours from observed-frame fluxes requires $K$-corrections and extinction corrections, which become increasingly important and galaxy-type dependent at higher redshift.

We use $g-r$ colour and $r$-band luminosity as proxies for star formation and stellar mass, respectively. Although these properties are correlated, the relationship is not deterministic: galaxies can be relatively faint but massive, or bright but less massive, depending on their stellar populations and star-formation histories. This variation means that colour provides information beyond luminosity alone, allowing the two tracers to produce distinct host-weighting distributions. It is these differences that enable us to constrain their relative contributions through the mixing fraction $\xi$. Since the simulated galaxy catalogue reproduces the observed correlations between galaxy properties, the extent to which these correlations make the two host-weighting prescriptions less distinguishable is naturally reflected in the resulting constraints on $\xi$.

A star formation-weighted prescription is most naturally associated with formation channels that preferentially trace actively star-forming stellar populations, such as compact binaries with relatively short delay times between formation and merger \citep{Artale_2019a}. Weights calculated with star formation weighting are denoted as $w_i'$ with a prime.

Our prior (Eq.~\ref{eq:event-catalogue-prior}) describes a mixture of the two weighting schemes. The mixing fraction $\xi$ controls the relative contribution of the two sub-populations to the overall event population. An intermediate value, $0<\xi<1$, can therefore be interpreted as a population containing events from both underlying formation scenarios, with $\xi$ setting their relative abundance and the limiting cases $\xi=1$ and $\xi=0$ corresponding to populations drawn entirely from one or the other.

\subsection{Approximations}
In a full hierarchical gravitational-wave analysis, the event-level likelihood is constructed from the strain data and therefore retains the information entering the detection statistic. Selection effects can then be treated consistently through the detection probability in the population normalisation. Our approximate marginal likelihood retains only the luminosity-distance and sky-position information and therefore does not contain sufficient information to reconstruct the signal-to-noise ratio for an arbitrary trial source. We consequently approximate the conditioning on detection by multiplying Eq.~\ref{eq:gwlikelihood} by
\begin{align}
\mathds{1}
\left[
\rho_k(z,H_0)>\rho_{\rm th}
\right],
\label{eq:detection-window}
\end{align}
which is equal to unity when the event would satisfy the detection threshold at a trial $(z,H_0)$, and zero otherwise. This restricts the reduced likelihood to regions of parameter space in which the event would have been detectable. In this way, selection effects are incorporated consistently within our approximate likelihood framework. The factor in Eq.~\ref{eq:detection-window} would not be required in an analysis using the full event-level likelihood.

\section{Simulation}
\label{subsec:sims}
In our simulation, the candidate host galaxies are drawn from the MICEcat mock galaxy catalogue \citep{Fosalba_2015}, a light-cone catalogue covering one-eighth of the sky out to $z=1.4$ and complete down to an $r$-band absolute magnitude of $M_r=-18.9$. The fiducial MICEcat cosmology is a flat $\Lambda$CDM model with $H_0=70~{\rm km\,s^{-1}\,Mpc^{-1}}$, $\Omega_m=0.25$, $\Omega_b=0.044$, and $\Omega_\Lambda=0.75$. Each galaxy is assigned a sky position, redshift, a true $r$-band absolute magnitude and $g-r$ colour. We neglect measurement uncertainty in the latter two properties. For computational efficiency, we restrict the MICEcat catalogue to
$z\leq0.7$. Since simulated hosts are selected from this catalogue, no mergers are generated beyond the same redshift. This truncation is therefore a modelling choice and should not be interpreted as the detection horizon of the network; an A\#-like detector would in practice
observe mergers at substantially higher redshift. For this analysis, we assume complete spectroscopic redshift coverage for all galaxies in the selected catalogue. This is an optimistic assumption; in a realistic application, redshift uncertainties \citep{crossparkin2025darksirensimpactredshift} and catalogue incompleteness \citep{Chen:2017rfc, Finke:2021aom,Gray_2022, Borghi_2026} would introduce additional modelling requirements and could weaken or bias the inferred constraints if not properly accounted for.

For each simulated event, we draw a host galaxy from the catalogue covering one eighth of the sky, according to the assumed host-weighting prescription, fixing the true sky position and redshift of the source. Binary parameters are then assigned, with the chirp mass drawn uniformly over $10\leq\mathcal{M}_c/M_\odot\leq50$ and the symmetric mass ratio uniformly over $0.10\leq\eta\leq0.24$, assuming a 100\% detector duty cycle. Gravitational-wave parameter-estimation uncertainties are approximated using the Fisher-matrix formalism implemented in \texttt{gwbench} \citep{Borhanian_2021}, yielding estimates of the luminosity-distance uncertainty and sky-localisation area. We model the gravitational-wave likelihood (Eq.~\ref{eq:gwlikelihood}) as a multivariate Gaussian in $d_L$, $\mathrm{RA}$, and $\cos(\mathrm{Dec})$.

The angular uncertainty is approximated as a two-dimensional Gaussian, with its width chosen such that 90\% of the probability is enclosed within the Fisher-estimated sky area. Candidate host galaxies are weighted by this angular likelihood when constructing the line-of-sight prior in Eq.~\ref{eq:event-catalogue-prior}, restricting the contribution to galaxies within four Gaussian widths of the estimated source position and thereby retaining more than 99.9\% of the localisation probability.

We simulate gravitational-wave observations with a three-detector network consisting of LIGO Hanford, LIGO Livingston, and Virgo. We consider two sensitivity scenarios: one in which the LIGO detectors operate at Advanced LIGO design sensitivity \citep{2015,Abbott_2020} and one in which they operate at the proposed A\# sensitivity. In each case, the Virgo detector is assigned the same noise power spectral density as the corresponding LIGO sensitivity curve, evaluated at the Virgo site. We adopt the Advanced LIGO zero-detuned, high-power design sensitivity \citep{Shoemaker2010} and the A\# target strain sensitivity \citep{KunsFritschel2024}, as implemented in \texttt{gwbench} \citep{Borhanian_2021}. 

These configurations are intended as idealised sensitivity scenarios rather than forecasts for a specific future observing run. In particular, the assumed detector duty cycle, simplified binary-mass distribution, and restriction of the source population to the $z\leq0.7$ MICEcat volume affect the distribution of signal-to-noise ratios and localisation uncertainties in the simulated sample. Consequently, the number of detected events considered here should not be interpreted as an estimate of the number of events obtainable within a particular observing time. Our aim is instead to assess how the host-population inference behaves for catalogues with different gravitational-wave measurement precision.

Events are retained only if their network signal-to-noise ratio satisfies
$\rho_{\rm net}>\rho_{\rm th}$, with $\rho_{\rm th}=12$. This detection model captures the leading dependence of detectability on luminosity distance, since more distant sources generally have lower signal-to-noise ratios and are therefore less likely to pass the detection threshold.

We estimate the selection factor, $p_{\rm det}$, in Eq.~\ref{eq:pdet} using a separate selection catalogue drawn from the same reference galaxy catalogue as the detected events. We use $N_{\rm sel}=5\times10^4$ injections, for which the effective sample size of the reweighted estimator exceeds $2\times10^4$ in every configuration, comfortably above the $4N_{\rm det}$ threshold beyond which Monte Carlo error in $p_{\rm det}$ can bias the inference \citep{Farr_2019}.
\section{Results}
\label{sec:results}


We consider catalogues containing $N_{\rm det}=200$ and $1000$ detected events for each of the Advanced LIGO and A\# sensitivity scenarios. We set the mixing parameter to $\xi=0.5$, which implies that half of the mergers occur in galaxies weighted by their star formation while the other half occur in galaxies weighted by their mass. 

\begin{figure}[t]
    \centering
    \includegraphics[width=\linewidth]{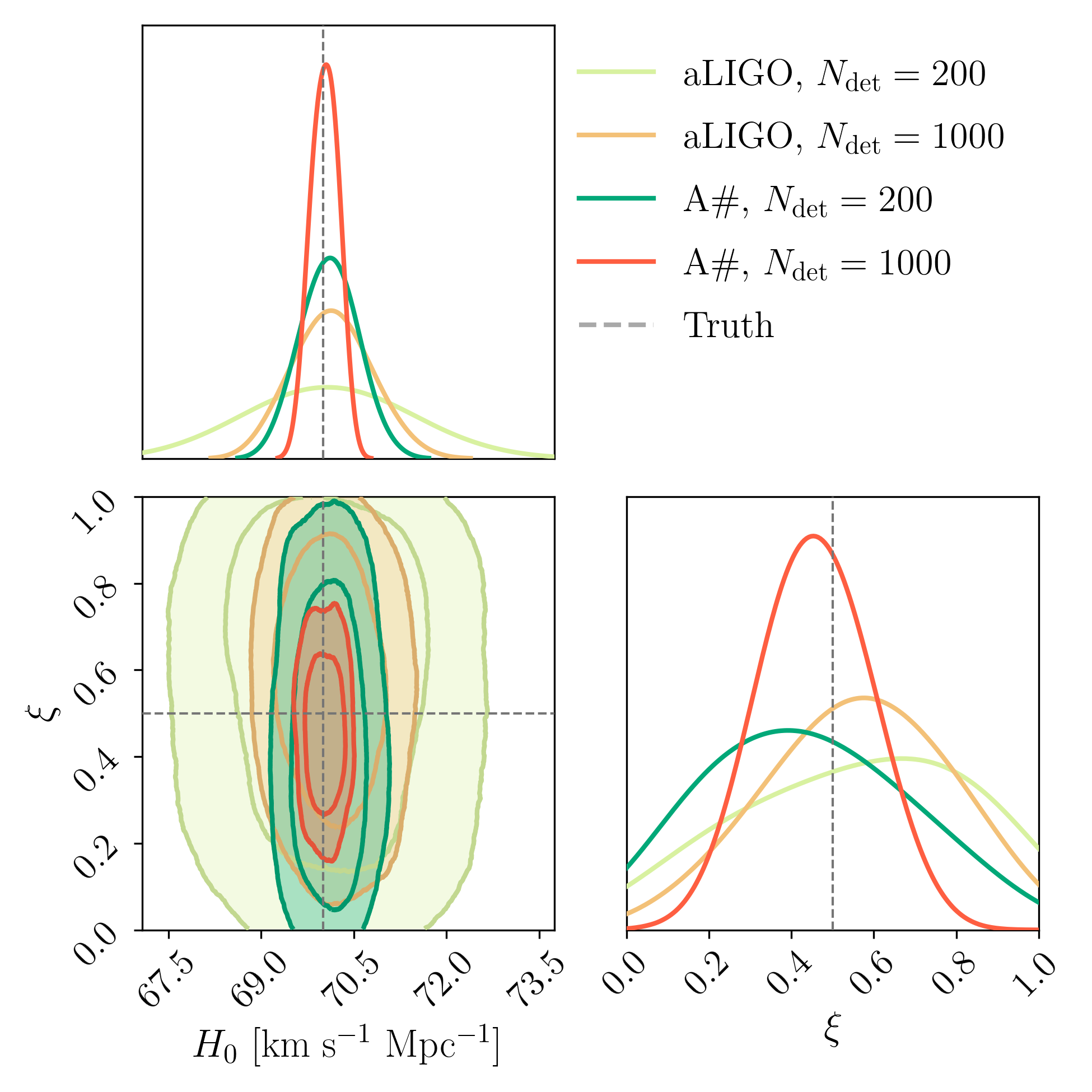}
    \caption{Posterior corner plot for $H_0$ and the mixing fraction, $\xi$, for different scenarios.
    The true values of $H_0, \xi$ are marked with dashed lines. The two-dimensional plot shows one- and two-sigma credible intervals.}
\label{fig:finalposteriors_ftrue05}
\end{figure}

We perform an MCMC exploration of the parameter subspace $\{H_0, \xi\}$ using the \texttt{emcee} ensemble sampler \citep{Foreman_Mackey_2013}. The contours in Fig.~\ref{fig:finalposteriors_ftrue05} and results in Table~\ref{tab:h0_f_constraints} show the average over 20 independent realisations. For each realisation, the individual-event likelihoods are multiplied to obtain the joint likelihood, from which the posterior is sampled. A two-dimensional KDE is then constructed from the posterior samples for each realisation.\footnote{For two hyperparameters, the likelihood can be evaluated directly on a grid, avoiding the uncertainty introduced by KDE. This becomes less feasible as the number of parameters increases.} We take the geometric mean of these posterior densities across the 20 realisations and sample the resulting distribution to obtain the contours and quoted $68\%$ credible intervals. The resulting geometric-mean posterior therefore represents the typical constraint expected from a single catalogue of $N_{\rm det}$ gravitational waves, while averaging over statistical fluctuations between independent realisations.

The averaged posteriors are consistent with the fiducial values of both $H_0$ and $\xi$ in all cases. Increasing the catalogue size from $N_{\rm det}=200$ to $N_{\rm det}=1000$ tightens the constraints for both detector networks. The $H_0$ uncertainty decreases by a factor of $\sim 1.9$ for A\# and $\sim 2.4$ for aLIGO, while the uncertainty on $\xi$ decreases by a more modest factor of $\sim 1.8$ for A\# and $\sim 1.3$ for aLIGO.

At fixed catalogue size, A\# provides substantially tighter constraints than aLIGO. For $H_0$, the A\# constraints are narrower by factors of $\sim 3.6$ and $\sim 2.8$ for $N_{\rm det}=200$ and $1000$, respectively. The improvement in $\xi$ is again more modest, with A\# reducing the uncertainty by factors of $\sim 1.3$ and $\sim 1.8$ for the same two catalogue sizes.

The joint constraints exhibit only a weak correlation between $H_0$ and $\xi$. As discussed further in Appendix~\ref{app:tracers}, this follows from the construction of the mock catalogue: neither host tracer correlates appreciably with redshift, so varying $\xi$ primarily redistributes host probability among galaxies within a redshift shell while leaving the overall host redshift distribution almost unchanged.

The mixing fraction becomes particularly informative for large A\# catalogues. For $N_{\rm det}\sim1000$, the posterior excludes $\xi=0$ and $\xi=1$ at $3.5\sigma$ and $3.8\sigma$, respectively. Such a result in real data would indicate that neither host-weighting prescription alone is sufficient to describe the merger population, providing evidence for multiple host subpopulations and potentially distinct binary black hole formation channels.

\begin{table}[t]
    \centering
    \renewcommand{\arraystretch}{1.4}
    \setlength{\tabcolsep}{10pt}
    \begin{tabular}{lccc}
    \toprule
    \textbf{Network} 
        & \textbf{$N_{\rm det}$} 
        & \textbf{$H_0\,[{\rm km\,s^{-1}\,Mpc^{-1}}]$} 
        & \textbf{$\xi$} \\
    \midrule
    aLIGO & 200
        & $70.04^{+1.20}_{-1.18}$
        & $0.59^{+0.27}_{-0.32}$ \\
    aLIGO & 1000
        & $70.13^{+0.57}_{-0.58}$
        & $0.57^{+0.21}_{-0.22}$ \\
    A\# & 200
        & $70.11^{+0.40}_{-0.44}$
        & $0.43^{+0.26}_{-0.24}$ \\
    A\# & 1000
        & $70.03^{+0.20}_{-0.22}$
        & $0.46^{+0.13}_{-0.12}$ \\
    \bottomrule
    \end{tabular}
    \caption{Averaged posterior constraints on $H_0$ and the host-population mixture parameter $\xi$ for the aLIGO and A\# detector networks. The quoted uncertainties correspond to the marginalised $68\%$ credible intervals, obtained as the geometric mean over 20 realisations. The injected values are $H_0=70\,\hcon$ and $\xi=0.5$.}
    \label{tab:h0_f_constraints}
\end{table}

\section{Comparison with recent work}
\label{sec: comparison}
\textcolor{black}{
Two papers closely related to this work were released while this analysis was in preparation. 
\citet{gray2026expandingscopedarksiren} introduce a framework in which the host-galaxy weighting is inferred jointly with cosmology. In their application, the host probability is described using a single luminosity tracer with a free power-law dependence, $p_{\rm host}\propto L^\epsilon$, where $\epsilon=0$ corresponds to uniform weighting and $\epsilon=1$ to weighting linearly with luminosity. They then perform a dark-siren analysis of the GWTC-5 data \citep{theligoscientificcollaboration2026gwtc50observationssecondfourth,theligoscientificcollaboration2026gwtc50constraintscosmicexpansion} using this parametrisation, allowing the data to determine how strongly the host probability depends on luminosity.}

\textcolor{black}{A similar luminosity-based parametrisation is adopted by \citet{agapito2026gravitationalwavedarksirensastrophysical}, who emphasise the astrophysical information contained in dark-siren host associations. Their framework jointly constrains the luminosity dependence of the merger probability and the probability that individual galaxies host a given event. Applied to GW170817 \citep{Abbott_2017a,Abbott_2017b}, they recover NGC~4993 as the most probable host. The three studies therefore have somewhat different emphases: \citet{gray2026expandingscopedarksiren} focus primarily on robust dark-siren cosmological inference, \citet{agapito2026gravitationalwavedarksirensastrophysical} on astrophysical host information and individual host identification, and our simulation-based analysis on whether a mixed host population can be distinguished well enough to reveal multiple merger subpopulations and, potentially, different formation environments.}

\section{Discussion and outlook}
\label{sec:disc}
Previous studies have pointed out that our choice of prescription for host galaxy weighting can create be a source of systematic error for dark siren measurements of $H_0$. Here we highlight that host weighting need not be viewed solely as a source of model misspecification---it also creates an exciting opportunity. 
By using a model that allows for multiple prescriptions, we can use patterns in the host association of binaries to glean astrophysical information about the formation pathways of binary black hole mergers. 

In particular, different galaxy properties encode different aspects of their star-formation histories: colour is sensitive to relatively recent star formation, whereas star formation rate more closely reflects the accumulated stellar population and may therefore favour galaxies whose stars formed at earlier epochs. 
By inferring the mixing fraction of different components, we can learn about whether mergers preferentially trace recent star formation or older stellar populations---all while guarding against misspecification. 

In this sense, modelling multiple types of host association can both avoid bias and provide a route towards probing compact binary formation environments. This interpretation, however, depends on the assumed host model: if additional formation channels contribute significantly, or if the merger rate depends on galaxy properties not captured by the tracers considered here, the inferred $\xi$ should instead be interpreted as an effective mixture within the adopted model. Future work should therefore extend the framework to include additional host properties and test the robustness of the inferred mixture fraction to unmodelled formation channels.

We estimate that a network of A\# observatories will be able to make astrophysically interesting statements about the mixing fraction of different types of host association after ${\cal O}(1000)$ events.

An important caveat is that these forecasts assume a complete galaxy catalogue. In practice, galaxy-catalogue incompleteness is itself correlated with many of the same properties that enter the host-weighting model. Incompleteness can therefore alter the observed distribution of these properties in a way that is potentially degenerate with the astrophysical weighting of the host population. For example, if low-luminosity or high-redshift galaxies are preferentially absent from the catalogue, an apparent preference for a luminosity- or mass-weighted host prescription could arise partly from property-dependent incompleteness rather than from the underlying binary formation channel. More generally, any incompleteness that affects the two weighting prescriptions differently can shift the inferred mixing fraction and thereby mimic differences in host association. This is particularly important if the mixing fraction is to be interpreted astrophysically, since an apparent preference for one host population could arise partly from catalogue incompleteness rather than from the binary formation channel itself.

The two-component model adopted here is necessarily simplified and assumes that the host dependence of the merger rate is adequately described by the two tracers considered. If additional formation channels contribute significantly, or if the merger rate depends on galaxy properties not captured by these tracers, such as metallicity, the inferred $\xi$ may no longer correspond directly to the underlying astrophysical branching fraction. Instead, it should be interpreted as an effective mixture within the adopted model. Extending the framework to include additional host properties and testing its robustness to model misspecification will therefore be important for future applications.

If more than one type of host association is identified in real data, the corresponding sub-populations would also be expected to exhibit different distributions of binary mass, mass ratio, effective spin, or spin-precession parameters. Isolated binary evolution and dynamical formation scenarios, for example, are expected to differ in both their preferred environments and their mass and spin distributions \citep{Mandel_2022}. This connection could be tested by dividing the gravitational-wave catalogue according to source properties or by extending the likelihood to model host weights and compact-binary population parameters simultaneously. In this way, host weighting would become not only a component of robust cosmological inference, but also a joint probe of binary black hole formation and evolution.

\begin{acknowledgments}
E.T. and P.D.L. are supported by Australian Research Council grants
LE210100002, DP230103088, and CE230100016. A. P. is supported by NSF Grant No. 2308193. This work used resources on the Vera Cluster
at the Pittsburgh Supercomputing Center (PSC). This work used Jetstream2 at Indiana University through allocation PHY240308 from the Advanced Cyberinfrastructure Coordination Ecosystem: Services \& Support (ACCESS) program, which is supported by National Science Foundation grants \#2138259, \#2138286, \#2138307, \#2137603, and \#2138296.
\end{acknowledgments}

\appendix

\section{Host-galaxy tracers in the mock catalogue}
\label{app:tracers}

The two host-weighting tracers introduced in Sec.~\ref{subsec:weights} are the percentile ranks of $r$-band luminosity and rest-frame colour. The mixing fraction $\xi$ is identifiable only to the extent that these two rankings order the catalogue differently, so it is useful to show how the two tracers are distributed in the mock catalogue.

\begin{figure}[t]
    \centering
        \includegraphics[width=0.9\linewidth]{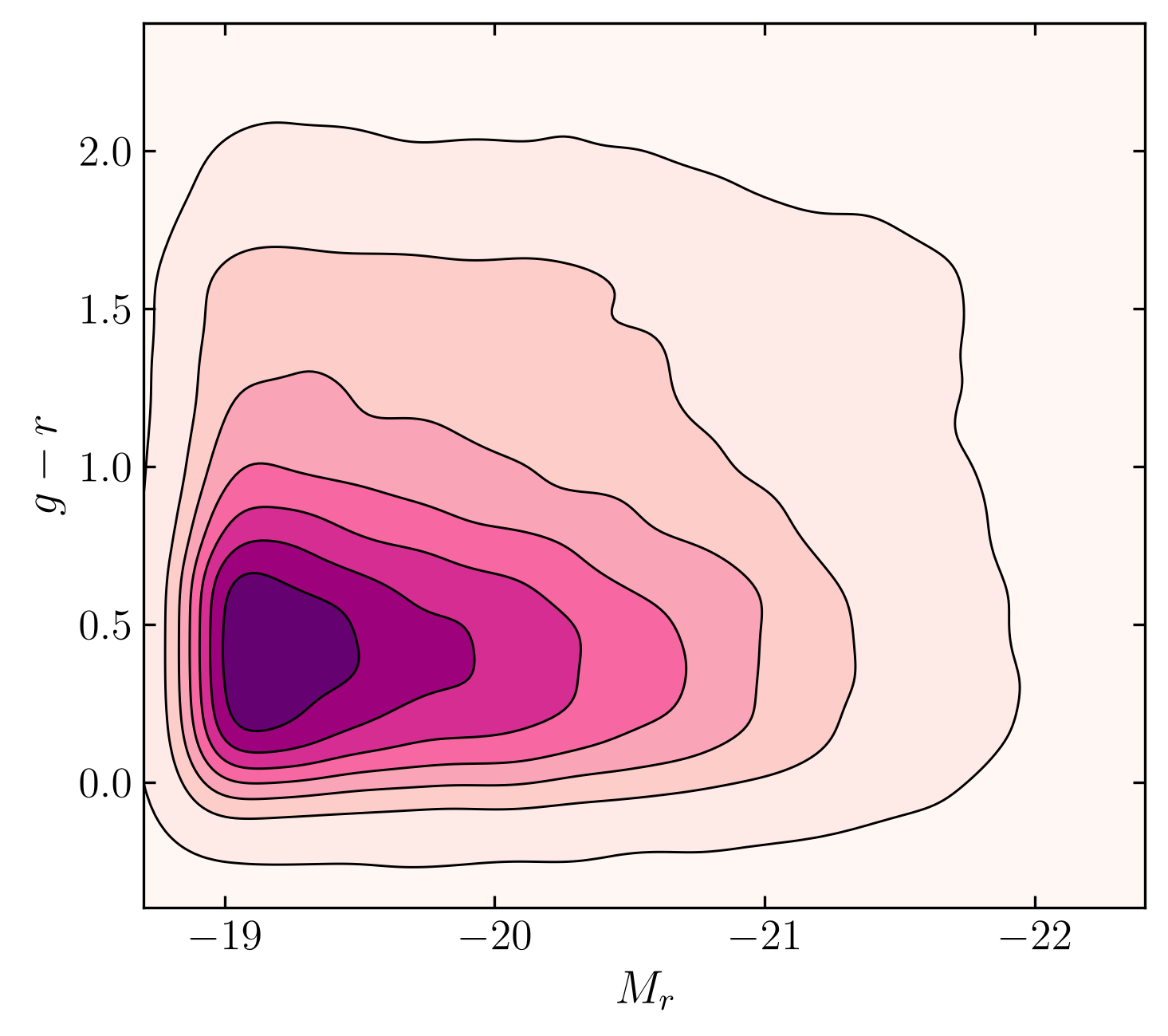}
        \caption{Joint distribution of $r$-band absolute magnitude $M_r$ and rest-frame colour $g-r$ for the MICEcat galaxies used in this analysis ($z\leq0.7$), shown as kernel-density contours. The two host-weighting tracers introduced in Sec.~\ref{subsec:weights} are the percentile ranks of these quantities. Their rank correlation is only $0.02$, and this near-independence allows the mixing fraction $\xi$ to be constrained.}
        \label{fig:contours}
\end{figure}

Fig.~\ref{fig:contours} shows the joint distribution of the two tracers for $1.5\times10^6$ MICEcat galaxies within the redshift range used in this analysis. Importantly, there is no strong diagonal correlation between them, meaning that a galaxy's luminosity rank carries almost no information about its colour rank. The two prescriptions therefore select genuinely different host populations and the data are able to distinguish between them. Had the two tracers been strongly correlated, every value of $\xi$ would have produced nearly the same host weighting, and $\xi$ would have been unconstrained regardless of catalogue size or detector sensitivity.

This near-independence is a property of the mock catalogue rather than of the method. Real magnitude-limited catalogues display a colour--magnitude relation, so the two rankings would be more strongly coupled and the constraint on $\xi$ correspondingly weaker. The forecasts of Sec.~\ref{sec:results} should be read with that caveat.

\bibliography{apssamp}

\end{document}